\documentclass[10pt, a4paper]{article}

\usepackage{lrec-coling2024} 
\usepackage{graphicx}
\usepackage{amsmath}

\title{Domain Adaptation for Dense Retrieval and Conversational Dense Retrieval through Self-Supervision by Meticulous Pseudo-Relevance Labeling}

\name{Minghan Li, Eric Gaussier} 

\address{Univ. Grenoble Alpes, CNRS, LIG \\
         Grenoble, France \\
         minghancs@163.com,  eric.gaussier@imag.fr\\}

\abstract{
Recent studies have demonstrated that the ability of dense retrieval models to generalize to target domains with different distributions is limited, which contrasts with the results obtained with interaction-based models. Prior attempts to mitigate this challenge involved leveraging adversarial learning and query generation approaches, but both approaches nevertheless resulted in limited improvements. In this paper, we propose to combine the query-generation approach with a self-supervision approach in which pseudo-relevance labels are automatically generated on the target domain. To accomplish this, a T5-3B model is utilized for pseudo-positive labeling, and meticulous hard negatives are chosen. We also apply this strategy on conversational dense retrieval model for conversational search. A similar pseudo-labeling approach is used, but with the addition of a query-rewriting module to rewrite conversational queries for subsequent labeling. This proposed approach enables a model's domain adaptation with real queries and documents from the target dataset. Experiments on standard dense retrieval and conversational dense retrieval models both demonstrate improvements on baseline models when they are fine-tuned on the pseudo-relevance labeled data.
 \\ \newline  }

\begin{document}

\maketitleabstract

\section{Introduction}
\label{Sec:intro}

Neural information retrieval (IR) has significantly improved IR systems through deep neural networks. It can be classified into two categories: interaction-based and representation-based (dense retrieval) approaches \citep{guo2020deep}. While interaction-based models generally outperform dense retrieval models, the latter are  preferred if one needs to deploy a model at large scale due to their speed advantage. However, recent studies, such as BEIR \citep{thakur2021beir}, have shown that dense retrieval (DR) models trained on a source domain generalize less well than traditional models as BM25 and interaction-based models on out-of-distribution (OOD) data sets. Training on target datasets with gold labels requires expensive annotations, posing limitations in real-world scenarios. Thus, addressing OOD scenarios for dense retrieval is crucial.

Domain adaptation aims to enable a model trained on a source domain to perform well on a target domain without using human labels \citep{wang2018DA,wang2022generalizing}. Several domain adaptation techniques have been proposed for dense retrieval. One approach is through data generation, as demonstrated by QGen \citep{ma2021zero}, which generates queries for the target domain using a query generator. However, the synthetic queries may not resemble real target queries. Another approach is domain adversarial learning \citep{wang2022generalizing}, exemplified by MoDIR \citep{xin2022zero}, which adversarially trains a dense retrieval encoder to learn domain-invariant representations. However, such a learning objective  may result in poor embedding spaces and unstable performance \citep{wang-etal-2022-gpl}.

In this paper, we address \textbf{do}main generalization for \textbf{d}ense
\textbf{re}trieval through \textbf{s}elf-\textbf{s}upervision by pseudo-relevance labeling (in short, DoDress).
We aim to build pseudo-relevance labels on the target domain using interaction-based models solely trained on the source domain, such as T5-3B \citep{nogueira2020document}, acting as re-rankers. This method eliminates the need for human annotations and allows the model to use genuine queries and documents from the target domain. Additionally, we investigate different negative sampling strategies \citep{zhou2022simans} to further enhance the final dense retrieval model on the target domain.

\begin{figure}
	\centering
	\includegraphics[width=200pt]{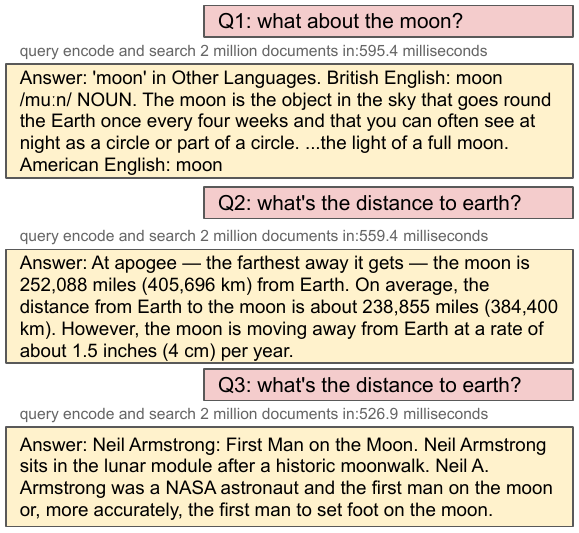}
	\caption{ {An example of conversational search (with our model deployed, top 1 as answer).}}
	\label{fig:convDR-example}
\end{figure}
Conversational search has become a prominent research area within information retrieval, involving natural conversations for information retrieval purposes \citep{zamani2023conversational,culpepper2018research}. Conversations exhibit contextualization, conciseness, and reliance on prior knowledge, presenting challenges for search systems in accurately understanding information needs.
{Figure~\ref{fig:convDR-example} shows an example of conversational search's query format: given a query, for example the third one, the system needs to understand the omission or user intention, by taking account into previous queries.
 Conversational dense retrieval (CDR) models have been developed to address this problem. However, CDR models require a large amount of training data and annotating relevance labels for target conversational search datasets is expensive.

To overcome these challenges, researchers have proposed approaches that leverage source domain data to mitigate data scarcity, such as \citep{yu2021few,lin2021contextualized}. However, these approaches do not utilize queries and documents from the target domain, leaving space for improvement when adapting to target domain data. In this paper, we propose a method that combines a query rewriting module with the pseudo-relevance approach for CDR models to alleviate the data and label scarcity issue of the target domain. This enables CDR models that are trained on a source domain like CANARD \citep{elgohary2019can} to better adaptation on the target domain using pseudo-relevance labels. 

Our contributions are threefold: First, we propose a pseudo-relevance labeling approach for a target IR dataset, which can be used to fine-tune a model to adapt better on the target domain. Second, we adopt the effective interaction-based model T5-3B trained on the source domain to generate pseudo-positive labels for the target domain. Besides, we explore different negative sampling strategies to enhance the final DR model. To the best of our knowledge, this is the first attempt to combine and investigate the two strategies. Third, we further apply the pseudo-relevance labeling approach to CDR models for conversational search by incorporating a query rewriting module, which is naturally a further step to apply the proposed pseudo-relevance labeling strategy.
This pseudo-relevance data complements CDR models and enables domain adaptation on the target dataset. Experimental results demonstrate the effectiveness of our approach, showing that fine-tuning DR models on the target pseudo-labeled data improves their performance, particularly benefiting the state-of-the-art approach GPL. Furthermore, further training CDR models on the generated training data from the target dataset leads to improved effectiveness \citep{yu2021few,lin2021contextualized}.

\section{Related Work}
\label{sec:append-how-prod}

\subsection{Pseudo-Quesries and Pseudo-Labeling for DR or IR}

QGen \cite{ma2021zero} proposes a generation approach for zero-shot learning in dense passage retrieval, using synthetic query generation.
Similarly, \citet{liang2020embedding} suggests using synthetic queries for unsupervised domain adaptation in dense passage retrieval. These papers highlight the effectiveness of query generation, which is also utilized in the GPL model \cite{wang-etal-2022-gpl}, leveraging a pre-trained T5 encoder-decoder \cite{raffel2020exploring}.
Recently, \citet{dai2022promptagator} prompt large language models (LLM) to create queries and train task-specific retrievers. However, they focus on the few-shot setting where a few annotated examples are required and do not focus on domain adaptation.
\citet{sun2021few} generate discriminative
queries based on contrastive documents.
Their approach also focus on the few-shot setting where a small volume of target data is required.
\citet{mokrii2021systematic} evaluate the transfer ability of BERT-based neural ranking models and use BM25 to generate pseudo-relevance labels. However, they don't focus on DR models and using only BM25 for pseudo-relevance labels may not be sufficient.
\citet{qu2021rocketqa} propose RocketQA which uses pseudo-labels for data augmentation, but this approach needs human label and does not focus on the domain adaptation of DR models.

\subsection{Conversational Dense Retrieval}
\label{secRelatedConv}

Conversational search presents unique challenges (see Section \ref{Sec:intro}).
Two commonly proposed approaches for addressing these challenges are query rewriting and conversational dense retrieval (CDR).
The query rewriting approach involves a module that rewrites conversational queries into a standard format for better handling by existing information retrieval systems \cite{mele2020topic,ren2018conversational,vakulenko2021question}.

The second approach is CDR with a query encoder to understand the conversational queries directly.  
\citet{mao2022convtrans} propose ConvTrans, that transforms web search sessions into conversational search sessions to address data scarcity of CDR. \citet{yu2021few} introduce ConvDR, a teacher-student framework that improves the few-shot ability of CDR by learning from a well-trained ad hoc dense retriever. CQE \cite{lin2021contextualized} uses annotated queries of the conversational query reformulation dataset CANARD \cite{elgohary2019can}  for the target datasets to train CDR. 
While these approaches still face domain gaps in the training data.

\section{Background}
\label{sec:appendix}

\paragraph{Dense retrieval} DR seeks to encode both queries and documents into a low-dimensional space with an encoder $g$, typically a BERT-like model \citet{karpukhin2020dense,xin2022zero}. The retrieval status value (RSV) of a query and a document is then calculated:
\begin{multline*}
    RSV(q,d)_{DR} = g(q) \cdot g(d)\  \,\, \\
    \left( or \,\, RSV(q,d)_{DR} = cos(g(q),g(d)) \right), \nonumber
\end{multline*}
%
where $g(q)$ (resp. $g(d)$) denotes the encoding of the query (resp. document).

\paragraph{BM25} BM25 \cite{RobertsonZ09} is a widely used standard IR algorithm based on term matching, without requiring to be trained:
\begin{multline*}
    RSV(q,d)_{BM25} = \\
    \sum_{w \in q \cap d} IDF(w) \cdot 
    \frac{tf_{w}}{k_1 \cdot (1-b+b\cdot \frac{l_{d}}{l_{avg}})+tf_{w}}, \nonumber
\end{multline*}
where $IDF(w)$ is the inverse document frequency, $l_{d}$ is the length of document $d$, $l_{avg}$ the average length of the documents in the data set, and $k_1$ and $b$ two hyper-parameters.

\paragraph{T53B} 
\citet{nogueira2020document} proposed to use T5 \cite{raffel2020exploring}
for IR by learning:
\begin{center}
\textit{Query: [q] Document: [d] Relevant: true or false}
\end{center}
\noindent where $[q]$ and $[d]$ are replaced with the query and document texts. During training, the T5 model learns to generate the word ``true'' when the document is relevant to the query, and the word ``false'' when it is not. 
The RSV is then determined by:
%
\begin{align*}
    RSV(q,d)_{T5}
    &= \frac{e^{Z_{true}}}{e^{Z_{true}}+e^{Z_{false}}}, \nonumber
\end{align*}
where $Z_{true}$ and $Z_{false}$ are the logits of output tokens.
\paragraph{Conversational Dense Retrieval }

The conversational dense retrieval (CDR) is similar to dense retrieval (DR), except the query format. A CDR architecture with pairwise learning is shown in Figure \ref{fig:tripletConvDr}, where the query encoder accepts the concatenation of conversational queries, to understand the user intention.

\section{Pseudo-Relevance Labeling for Dense Retrieval}
\label{Sec:pseudo_dense}

To enhance the domain generalization ability of DR models, we employ the pseudo-relevance labeling strategy. This involves finding pseudo-positive and pseudo-negative documents for a target dataset's queries, which are used to fine-tune the DR models. Our approach includes BM25 hard negative sampling (Figure~\ref{figArchBMHard}) and our best approach using SimANS hard negative sampling (Figure~\ref{figArch}). We will discuss them further below.

\begin{figure}[htbp]
\centering
\includegraphics[width=210pt]{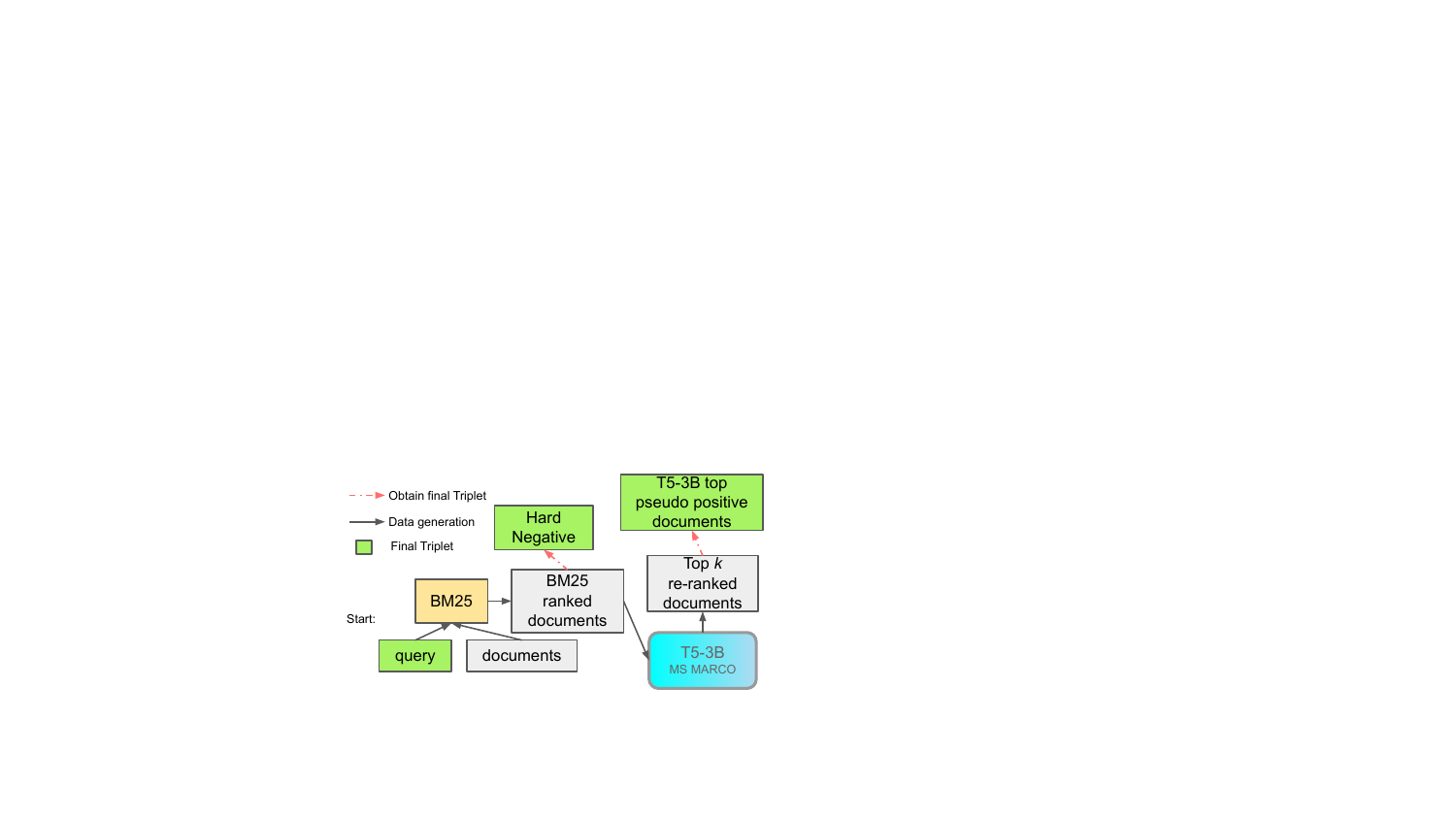}
\caption{The overall pipeline with BM25 hard negative sampling for pseudo-relevance labeling.} \label{figArchBMHard}
\end{figure}

\begin{figure}[htbp]
\centering
\includegraphics[width=220pt]{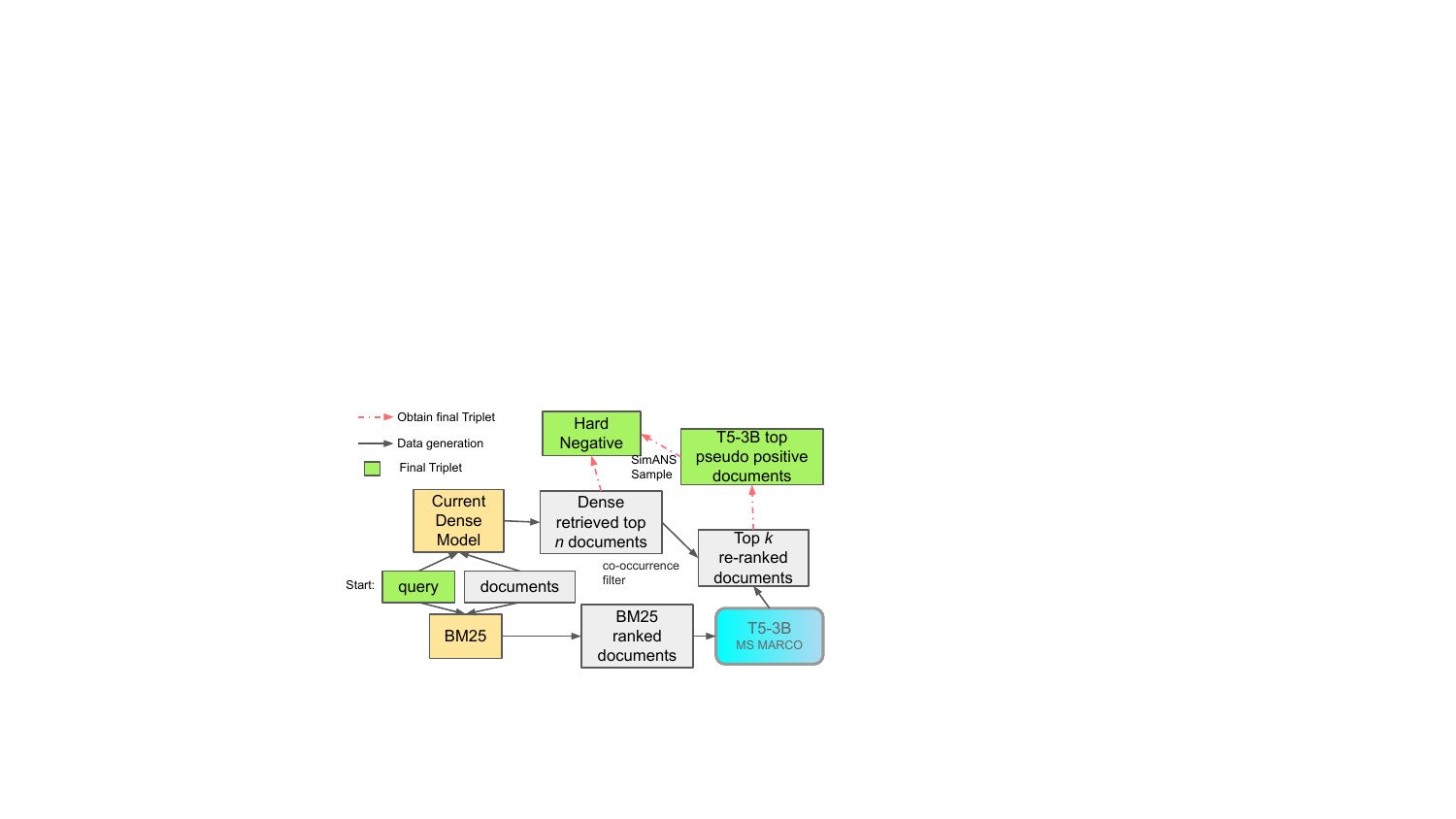}
\caption{The overall pipeline of generating self-supervised data with meticulous pseudo-relevance labeling using SimANS hard negative sampling.} \label{figArch}
\end{figure}

\subsection{Pseudo-Positive Sampling} \label{sec:positive}
We propose here to consider, for each query, the top $k$ documents obtained with the combination BM25\&T53B, in which T5-3B serves as a re-ranker, as relevant (or positive). $k$ is a hyper-parameter which can be set according to different information, as, \textit{e.g.}, the number of available queries and documents. T5-3B, which has been shown to be a good zero-shot IR model in \cite{nogueira2020document}, is fine-tuned on MS MARCO collection.

\subsection{Pseudo-Negative Sampling}
Furthermore, for each relevant query-document pair, we sample $m$ documents and consider them as non-relevant (or negative). Different negative mining strategies can be used, as described below.
 For each query, $k\times m$ query-document triplets (query, positive document, negative document) can be formed. The green blocks in Figure~\ref{figArchBMHard} and Figure~\ref{figArch} represent the elements constituting these triplets. 

\paragraph{Global and BM25 Hard Negative Sampling}
A simple negative mining strategy is global random negative sampling which consists in sampling, from all non-positive documents in the dataset, $m$ documents which are considered as negative.

A key challenge in DR is to construct proper negative instances for learning its representations \cite{karpukhin2020dense}. Previous global random negative instances might be too simple for the DR models.
So, for each query, we further propose to use the BM25 top ranking documents, again excluding the positive documents, as hard negative instances for training the DR models. The architecture is shown in Figure \ref{figArchBMHard}. 

\paragraph{Meticulous Hard Negative Sampling}

Recently, SimANS \cite{zhou2022simans} shows existing negative sampling strategies \cite{karpukhin2020dense,xiong2020approximate} suffer from the uninformative
or false negative problem, and the authors show that the negatives ranked around the positives 
are generally more informative and less likely to be false negatives.
This leads to a sampling probability distribution of the form \cite{zhou2022simans}:
\begin{equation}
\small
\label{eq-simans}
p_{i} \propto \exp{(-a(s(q,d_{i})-s(q,\tilde{d}^{+})-b)^{2})}, 
\forall~d_{i} \in \widetilde{\mathcal{D}}^{-},  
\end{equation}
where $a$ controls the density of the distribution, $b$ controls the peak of the distribution, $\tilde{d}^{+}\in \mathcal{D}^{+}$ is a randomly sampled positive, and
$\widetilde{\mathcal{D}}^{-}$ is the top-$k$ ranked negatives.

In this paper, we use this SimANS approach with the positive documents obtained in Section~\ref{sec:positive}.
The architecture is shown in Figure~\ref{figArch}: we select hard negatives that are around the positive instances in the top ranking of current DR models (i.e., D-BERT and GPL respectively), thus more ambiguous and informative negatives can be sampled. 
\ref{append:simansDetail}. 
The green blocks in Figure~\ref{figArch} correspond to the queries and the associated positive and hard negative documents.

\subsection{Improving GPL: Combining Pseudo-Relevance Labels and Pseudo-Queries}
\label{baseApproachSec}
To enhance both the QGen and GPL approaches which rely on pseudo-queries, we suggest further training such models like GPL using the proposed pseudo-relevance triplets. We believe one can gain from this additional training on the target collection as pseudo-queries and pseudo-relevance labels rely on different sources of information and are complementary to each other. In our experiments, we demonstrate that this combination significantly improves the pseudo-query generation approach.

\section{Pseudo-Relevance Labeling for Conversational Dense Retrieval}
\begin{figure}
	\centering
	\includegraphics[width=160pt]{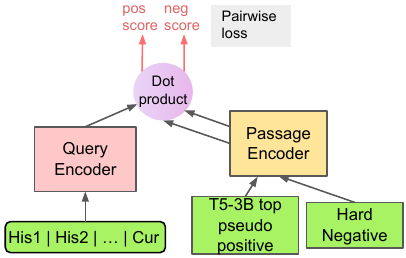}
	\caption{CDR architecture with training.}
	\label{fig:tripletConvDr}
\end{figure}

The architecture for training the CDR model with pairwise loss is shown in Figure \ref{fig:tripletConvDr}, which is similar to the DR model except with a different query format. 
CDR models face data scarcity issues and have potential for improvement through domain adaptation. Naturally, the pseudo-relevance labeling approach  can be used to help the CDR models by modifying the previous section slightly.

To train the CDR model, we concatenate the history and current queries, aiming to teach the query encoder to generate a de-contextualized query representation. Additionally, we need annotations for positive and negative documents. Our solution is pseudo-relevance labeling with conversational queries.

To achieve this, we train a T5-Large sequence-to-sequence model on CANARD \cite{elgohary2019can}, which is a dataset for learning to rewrite conversational queries. 
The overall architecture for the proposed approach in this section is shown in Figure~\ref{fig:convDR} and the detailed procedures are outlined below.

\begin{figure}[htbp]
	\centering
	\includegraphics[width=220pt]{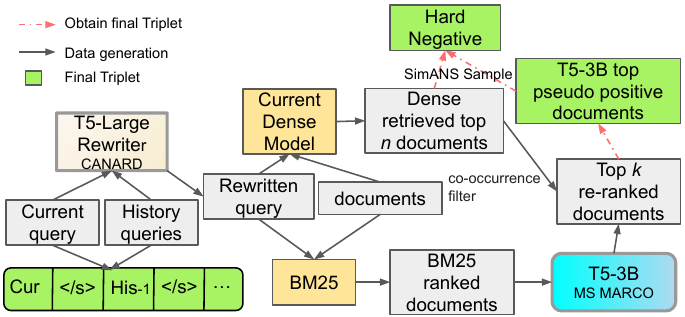}
	\caption{Overall pipeline of generating pseudo-data for conversational dense retrieval. 
 }
	\label{fig:convDR}
\end{figure}

\subsection{T5-Large Query Rewriter Module}
This module is shown in the left part of Figure~\ref{fig:convDR}.
We utilize the T5 model's special token "</s>" to concatenate the current query and history queries. The current query is placed at the beginning, followed by the recent history queries (with farther history queries located towards the end), following a similar approach as described in \cite{mao2022convtrans}:
\begin{center}
		$</s> cur </s> his_{-1} </s> his_{-2}...$
 \end{center}
The T5-Large model is trained using CANARD \cite{elgohary2019can}  with ground-truth human rewritten queries serving as labels, to generate rewritten queries that comprehensively capture the user intentions. 
The T5-Large query rewritten model can effectively rewrite conversational queries for target datasets into the desired de-contextualized queries. Then with them, we can sample pseudo-positive and pseudo-negative documents for them, as illustrated in Figure~\ref{fig:convDR}.

\subsection{Pseudo-Data Format of Conversational Dense Retrieval Model}
\label{Sec:trainCDR}
The sampling step is similar to the approach shown in Section \ref{Sec:pseudo_dense}: with T5-3B and SimANS for the generation of pseudo-labels given the 
rewritten queries.

Consequently, for training a CDR model as shown in Figure \ref{fig:tripletConvDr},
each line of the training triplet file can be presented in the following format:
\begin{center}
	ConcatenateQ $\backslash$t Texts of a Pseudo-Positive Doc $\backslash$t Texts of a Pseudo-Negative Doc
\end{center}
where $ConcatenateQ$ follows the format in \cite{lin2021contextualized}:
\begin{center}
	$hisQ_{1}| hisQ_{2}|...|curQ$
\end{center}
The triplet training file can serve as training data for finetuning a CDR model from a source domain. If the generated data is large enough, it even may be used to train a query encoder from scratch.

\section{Experiments}
We conducted experiments on both DR and CDR models for domain adaptation based on the approaches described in the previous sections.  In the remainder of this section, we first describe the setup of DR experiments and analyze their results; we then detail the CDR experiments.

\subsection{Domain Adaptation for Dense Retrieval Experimental Setup}
\label{sec:exp_domain_dense}
\paragraph{Datasets}
The MS MARCO passage ranking data set ~\cite{bajaj2018ms} is used as the source domain data. We want to experiment in an extreme scenario where no test queries can be seen during training even without human labels. This is to say, we need to generate the pseudo training data with the training queries which is not in the test set. To do so,
we experiment on 3 target domain data sets from the BEIR benchmark \cite{thakur2021beir}. They are FiQA, finance question answering \cite{fiqa-2018} which contains 6000 training queries, 
BioASQ biomedical question answering \cite{bioasq-2015} (following \cite{wang-etal-2022-gpl}, irrelevant documents are randomly eliminated, leaving 1M documents) which contains 3243 training queries from original collection\footnote{\url{http://participants-area.bioasq.org/Tasks/8b/trainingDataset/}}, 
and Robust04, news documents \cite{robust04-2005} which contains 250 queries. Different topics and tasks are covered by these chosen data sets. For Robust04, we select the first 100 queries as training and development set, and the last 150 queries are used as test set.

\paragraph{Experimental Setting}
\label{Sec:DR_exp_set}
The DR model, called D-BERT, is based on the DistilBERT \cite{DBLP:journals/corr/abs-1910-01108} with 6 layers. D-BERT is initially trained on the source domain. Two groups of experiments are conducted: one with D-BERT alone and the other with 
the GPL model \cite{wang-etal-2022-gpl} as start points.
Both models are trained using the RankNet pairwise loss  \cite{burges2010ranknet,li2022bert} on the generated triplets.

Table \ref{tab.stat} provides details on the number of queries, the selected value of $k$ (top documents as relevant), and the number $m$ of non-relevant documents per relevant document for each dataset.
A development set is created to select hyperparameters, consisting of 10 relevant documents and 90 randomly selected non-relevant documents per query. The best model is saved based on the NDCG@10 score on the development set.

A maximum sequence length of 350 with mean pooling and dot-product similarity is used. A batch size of 8 and a learning rate of 2e-6 with Adam optimizer are employed for 10,000 training steps. Cosine learning rate decay \cite{loshchilov2017sgdr} is utilized. For SimANS, the hyperparameters $a$ and $b$ are set to 0.5 and 0, respectively.

\begin{table}[htbp]
\caption{The top $k$ selected as positive and $m$ as negative for each data set. The number in parentheses is used for generating training data, remaining for dev set. }\label{tab.stat}
\centering
\scalebox{0.7}{
\begin{tabular}{|c|c|c|c|c|c|}
\hline
data set &  \#queries (exclude test) & \#docs  & $k$ & $m$\\
\hline
FiQA & 6000 (5960) & 57K    &1 & 10\\
BioASQ & 3243 (3193) &1M& 2 & 15\\
Robust04 & 100 (90) &528K&  15 & 67\\
CAsT-19 &269 (219) &2M &5 &100\\
\hline
\end{tabular}}
\end{table}

\paragraph{Baselines}
We compare our proposed approaches with various existing methods in the field, including:
\begin{itemize}
    \item Zero-shot models: BM25 based on Anserini \cite{yang2018anserini} and D-BERT trained solely on the source collection.
    \item Pre-training based models: we use three state-of-the-art models, namely SimCSE \cite{gao2021simcse}, ICT \cite{lee2019latent}, and TSDAE \cite{wang2021tsdae}.
    \item Domain adaptation approaches: MoDIR \cite{xin2022zero}, UDALM \cite{karouzos2021udalm}, QGen \cite{ma2021zero}, and GPL \cite{wang-etal-2022-gpl}. The combination of GPL with TSDAE is currently considered as the best approach.
\end{itemize}

Furthermore, we include interaction-based models BM25+CE and BM25+T53B as strong baselines, which re-rank the top 100 BM25 ranked list using \textit{ms-marco-MiniLM-L-6-v2} and T5-3B cross encoders\footnote{\url{https://huggingface.co/castorini/monot5-3b-msmarco} which is trained on MS MARCO for 100K steps.}, respectively. In this paper, this T5-3B model is also used for pseudo-positive labeling.
These models are known for their good performance in out-of-domain settings \cite{thakur2021beir}.

\paragraph{Results and Analysis}

Table \ref{tab.fiqa} displays the results obtained with different models and approaches. The results reported for BM25+CE, UDALM, MoDIR, SimCSE, ICT, TDSAE, QGen and TSDAE+GPL are from \cite{wang-etal-2022-gpl}. 
Since we test the Robust04 on the last 150 queries, for BM25+CE, GPL and TSDAE+GPL, we load the trained checkpoints of D-BERT from \cite{wang-etal-2022-gpl}\footnote{\url{https://huggingface.co/GPL}}, and evaluate them on the last 150 queries. 
The notation ``DoDress-T53B (D-BERT)'' corresponds to the D-BERT dense retrieval model pre-trained on MS MARCO and fine-tuned on the target data using the pseudo-relevance labels generated with BM25+T53B. 
The notation (GPL) means the same for GPL, which is first trained on the target pseudo queries it generates and associated documents prior to be trained on the target triplets.

We address three main research questions, denoted as \textbf{RQ}.

\begin{itemize}
    \item[\textbf{RQ1}] Do BM25+T53B top positives help domain generalization for dense retrieval models?
\end{itemize}
The results in Table \ref{tab.fiqa} shows that DoDress-T53B (D-BERT) and DoDress-T53B (GPL) outperform D-BERT and GPL, respectively, on the FiQA and Robust04 datasets using different negative sampling strategies. Notably, DoDress-T53B (D-BERT) with the SimANS negative mining strategy achieves an 11.5\% improvement over D-BERT, while DoDress-T53B (GPL) shows an 8.6\% improvement over GPL. On the BioASQ dataset, the approach with global random negative sampling fails, but the proposed approach with the other two negative sampling strategies improves D-BERT and GPL, respectively. 
The reasons of this success can be explained by the fact that the proposed pseudo-relevance labeling approach enables the DR models to see and be trained with real queries and documents of the target dataset. This labeling approach is further improved when combined with query generation approach of GPL. 
These results demonstrate that the proposed pseudo-relevance labeling approach helps dense retrieval models generalize to new domains. As the reader may have noticed, the choice of the negative sampling strategy is crucial for its effectiveness.

\begin{itemize}
    \item[\textbf{RQ2}] What is the best negative sampling strategy?
\end{itemize}
From Table \ref{tab.fiqa}, we observe an overall ascending trend in performance with the three different negative sampling strategies. The global random negative method shows improvements on FiQA and Robust04 but fails on the BioASQ dataset. This may be due to uninformative negatives that are too easy for the dense retrieval models on the target domain. In contrast, the BM25 hard negative and SimANS negative sampling strategies outperform the global random negative strategy, improving D-BERT and GPL on all three datasets. This highlights the importance of sampling hard negatives in the proposed pseudo-relevance labeling data generation approach. 
Among the three strategies, SimANS hard negative sampling consistently performs the best on all datasets, surpassing the global random negative and BM25 hard negative strategies. 

\begin{itemize}
    \item[\textbf{RQ3}] What is the best overall approach?
\end{itemize}
As one can note from Table~\ref{tab.fiqa}, the DoDress-T53B models outperform all other models on all collections but BM25 on BioASQ and of course the models consisting in re-ranking BM25 results with cross-encoders, which constitute an upper bound and are too costly to be used in practice. In addition, if BM25 is a strong competitor for domain adaptation, as reported in \cite{thakur2021beir}, its performance vary significantly from one collection to the other (very good on BioASQ, very poor on FiQA). Lastly, it is interesting to note that DoDress-T53B (GPL) outperforms the previous state-of-the-art model (TDSAE+GPL) by a large margin on BioASQ and Robust04, showing the effectiveness of the proposed approach.

\begin{table}[htbp]
\caption{Domain adaptation result of FiQA, BioASQ and Robust04 (during training only use train queries).}\label{tab.fiqa}
\centering
\scalebox{0.7}{
\begin{tabular}{|l|c|c|c|c|} 
\hline
\textbf{Method}
& \textbf{FiQA}          & \textbf{BioASQ}       & \textbf{Robust04}  & \textbf{Avg.} \\
\hline
\multicolumn{5}{|l|}{\textit{Zero-Shot Models}} \\
\hline
D-BERT & 26.7 &53.6 &39.1 &39.8 \\
BM25 (Anserini) & 23.6 &73.0&44.4 & 47.0\\
\hline
\multicolumn{5}{|l|}{\textit{Re-Ranking with Cross-Encoders (Upper Bound)}} \\ \hline
BM25 + CE   & 33.1 &72.8&45.8&50.6\\
BM25 + T53B   & 39.2&76.1&51.8&55.7\\ \hline
\multicolumn{5}{|l|}{\textit{Previous Domain Adaptation Methods}} \\ \hline
UDALM                                      & 23.3 &33.1&- &-\\
MoDIR (ANCE) & 29.6&47.9&-&-\\
\hline
\multicolumn{5}{|l|}{\textit{Pre-Training based: Target $\to$ D-BERT}}                                       \\ 
\hline
SimCSE & 26.7&53.2&-&- \\
ICT & 27.0&55.3&-&- \\
TSDAE & 29.3&55.5&- &-\\
\hline
\multicolumn{5}{|l|}{\textit{Generation-based (Previous SOTA)}} \\ 
\hline
QGen    & 28.7&56.5&-&-\\
GPL & 32.8&62.8&41.9 &45.8\\
TSDAE + GPL &  34.4&61.6&40.7 &45.6\\
\hline

\multicolumn{5}{|l|}{Proposed: T53B, Global Random Neg}
  \\ 
\hline
DoDress-T53B (D-BERT) &27.3&52.9&40.5&40.2

\\
DoDress-T53B (GPL) & 33.0&62.0&43.2&46.1

\\ \hline
\multicolumn{5}{|l|}{Proposed: T53B, BM25 Hard Neg}       \\ 
\hline
DoDress-T53B (D-BERT) & 30.4&58.6&41.6&43.5

\\
DoDress-T53B (GPL) & 34.2&64.7&43.3&47.4

\\
\hline
\multicolumn{5}{|l|}{Proposed: T53B, SimANS Hard Neg}
  \\ 
\hline
DoDress-T53B (D-BERT) & \textbf{31.0} & \textbf{60.6}& \textbf{43.6} & \textbf{45.1}

\\ 
DoDress-T53B (GPL) & \textbf{34.9}& \textbf{65.3}& \textbf{45.5} & \textbf{48.6}

\\ \hline
\end{tabular}}
\end{table}

\subsection{Conversational Dense Retrieval Experimental Setup}

\paragraph{Dataset Used}
We utilize the TREC CAsT 2019 (CAsT-19) dataset \cite{dalton2020trec}. 
CAsT-19 comprises 30 training topics and 20 test topics, with each topic representing a conversational search session consisting of queries from multiple turns. The dataset contains a total of 269 training queries.
Notably, in this paper, we investigate an extreme scenario in which we do not have access to human rewritten queries and relevance labels for CAsT-19, resulting in an almost zero-shot scenario.

For efficient experiments, we furthermore follow the experimental protocol defined in \cite{wang-etal-2022-gpl}: we randomly remove irrelevant passages from the whole 38M TREC CAsT-19 corpus to obtain a smaller corpus consisting of 2M passages. 

\paragraph{T5 Rewriter}

The conversational rewriter used is the T5-Large version\footnote{\url{https://huggingface.co/t5-large}}. We train the model on CANARD dataset \cite{elgohary2019can} which contains 31526 training instances repeatedly for 80K instances, and evaluate the T5-Large model on development set for every 20000 instances and save the best model. 
The learning rate is 5e-5 and batch size is 4 using AdamW optimizer \cite{loshchilov2017decoupled}. 
The BLEU \cite{papineni2002bleu} results of the final T5-Large model on CANARD are presented in Table \ref{tab.canard}. 
We can see the T5-Large model can obtain near human accuracy for rewriting conversational queries on CANARD dataset. 
Besides, we have computed its BLEU score on TREC CAsT-19, obtaining 64.35
, which indicates a favorable outcome.

\begin{table}
	\caption{BLEU scores of different approaches for rewriting conversational queries on CANARD dataset. 
 The first four methods are baseline approaches used in \cite{elgohary2019can}.}\label{tab.canard}
	\centering
	\scalebox{0.7}{
		\begin{tabular}{l c c}
			\hline
			Method &  Dev & Test  \\
			\hline
			Copy & 33.84 & 36.25   \\ \hline
			Pronoun Sub & 47.72 & 47.44\\ \hline
			Seq2Seq &51.37 & 49.67 \\ \hline
			Human Rewrites & \multicolumn{2}{c}{59.92}\\ \hline
			T5-Large & 59.5 & 57.9\\\hline
		\end{tabular}
	}
\end{table}

\paragraph{Baselines}

\begin{itemize}
    \item Zero-Shot baselines: 
The BERT-dot-v5\footnote{\url{https://huggingface.co/sentence-transformers/msmarco-bert-base-dot-v5}} model trained on MS MARCO is used as zero-shot baselines. We experiment with four methods:

(1) using only the current query, (2) concatenating the history and current queries (it should be noted that the query encoder is not specifically trained to handle this format), (3) using T5-Large rewritten queries for retrieval (may not be efficient in real-world scenarios), and (4) the upper bound method with human rewritten queries from the dataset. 
In addition, we also include BM25 using Anserini \cite{yang2018anserini} with official rewritten queries and T5-Large rewritten queries as baselines.

\item Cross-Encoder: We adopt T5-3B model which is trained on MS MARCO (same version as before) to rerank the BM25 list from T5 Rewritten queries. 
\item Related work: ConvDR \cite{yu2021few} and CQE \cite{lin2021contextualized} are compared, which are CDR models aiming to address the data scarcity issue.
They are trained from the BERT-dot-v5 checkpoint. We train ConvDR for 40k steps with the batch size of 8. We observed that the performance remained similar across different training intervals, namely 10k, 40k, and 80k steps. 
Following \cite{lin2021contextualized}, we train CQE for 120k steps with a batch size of 8, which is comparable to their original paper's training process of 20k steps with a batch size of 96. The learning rate used in our training process is set to 2e-6.
\end{itemize}

\paragraph{CDR Training}
We further conduct experiments based on the learned checkpoints of baseline models ConvDR and CQE to deal with the few-shot learning scenario. We fine-tune them with our generated pseudo-labels.
Following \cite{yu2021few,lin2021contextualized}, the passage encoder is fixed and the query encoder is trained. All history turns are used since they are short.
The parameters for generating pseudo-relevance labels are the same as before (Table \ref{tab.stat}).  
Our fine-tuning strategy is similar to the previous DR experiments, using the RankNet pairwise loss with a batch size of 8 and learning rate of 2e-6. We train on our (few shot) pseudo-target data for 2000 steps and evaluate on the development set every 500 steps, saving the best models. Finally, we report the results of the trained models using NDCG@3, which is consistent with previous research \cite{yu2021few,lin2021contextualized}.

\paragraph{Experiment Result}

Experiment results are presented in Table \ref{tab.cast19}. Besides, we also show the deployed demo of our best trained CDR model (58.0) dealing with real world conversational queries in Table \ref{fig:convDR-example} (deployed on a RTX 6000 GPU).
\begin{table}[htbp]
	\caption{Domain adaptation result of CAsT-19.}\label{tab.cast19}
	\centering
	\scalebox{0.7}{
		\begin{tabular}{|c|c|}
\hline
model &  nDCG@3 (\%) \\
\hline
\multicolumn{2}{|c|}{\textit{Zero-Shot Models}} \\
\hline

BERT-dot-v5(current) & 33.4 \\ 
BERT-dot-v5(concatenation) & 27.2 \\ 
BERT-dot-v5(T5Rewrite) & 53.2 \\ 
BERT-dot-v5(Human) (Upper Bound) & 58.9 \\ 
BM25(Human) & 37.0\\
BM25(T5Rewrite) &31.2\\
\hline
\multicolumn{2}{|c|}{\textit{Re-Ranking with Cross-Encoders}} \\ \hline
T5-3B rerank T5Rewrite   & 56.7\\ \hline

\multicolumn{2}{|c|}{Related Work}\\
\hline
ConvDR (BERT-dot-v5) & 55.4

\\
CQE (BERT-dot-v5) & 53.7

\\ \hline
\multicolumn{2}{|c|}{Proposed Approach}\\
\hline

\multicolumn{2}{|c|}{T53B, SimANS Neg, based on ConvDR}
  \\ 
\hline
DoDress-T53B (BERT-dot-v5) & \textbf{58.0}

\\ \hline
\multicolumn{2}{|c|}{T53B, SimANS Neg, based on CQE}
  \\ 
\hline
DoDress-T53B (BERT-dot-v5) & \textbf{57.6}

\\ \hline
\end{tabular}}
\end{table}
We address here two main research questions.
\begin{itemize}
    \item[\textbf{RQ1}] In the context of conversational search, how do IR models exclusively trained on MS MARCO perform as zero-shot models?
\end{itemize}

The best baseline among zero-shot models is BERT-dot-v5(Human), which constitutes an upper bound as it uses human rewritten queries on the target dataset. This model largely outperforms BM25(Human). In comparison, using the current query or the concatenation of queries with the BERT-dot-v5 model yields lower results than the human rewritten queries, respectively 33.4 and 27.2 compared to 58.9. 
This means that, for this dataset, although DR models can successfully retrieve documents when given real human rewritten queries and outperform BM25 approach, they do not perform well when they are not specifically fine-tuned to understand the conversational queries or when they are not given good rewritten queries. When using T5 rewritten queries, the performance becomes closer to the one obtained with human rewritten queries. In addition, using T5-3B with the T5 rewritten queries as a reranking model yields very good results, close to the upper bound.

\begin{itemize}
    \item[\textbf{RQ2}] Does the proposed in-domain pseudo-relevance data generation approach effectively enhance the performance of CDR models?
\end{itemize}

Firstly, let's discuss related works on CDR. ConvDR and CQE achieve scores of 55.4 and 53.7, respectively, 
outperforming standard zero-shot DR model baselines except the upper bound using human rewritten queries. These models benefit from the CANARD dataset, which enables effective representations for conversational queries. However, as the queries in CANARD differ from the ones in the target dataset, training with in-domain queries becomes crucial.  
The proposed models based on ConvDR and CQE achieve scores of 58.0 and 57.6, respectively, representing improvements of 4.7\% and 7.3\% over ConvDR and CQE.  
This is due to the fact that the proposed pseudo-relevance labeling approach enables the CDR models to see real queries and respective documents on the target domain, resulting in better adaptation.
Overall, our proposed approach achieves comparable performance to the one obtained by the BERT-dot-v5 model with real human rewritten queries, without requiring human annotations on the target dataset.


\section{Conclusion}
 This paper first studied whether one can benefit from existing re-ranking based IR models, pre-trained on MS MARCO, to generate pseudo-relevance labels for an unannotated, target collection. These labels, along with sampled positives and negatives, are used to fine-tune dense retrieval models on the target collection. The experiments revealed that carefully generating pseudo-labels improves the generalization results of DR models and that additional improvements can be obtained with the query generation approach of GPL. 
 We also investigated several negative sampling strategies, based on BM25 and SimANS, and confirmed the importance of identifying useful hard negative documents. 
 The proposed pseudo-relevance labeling approach has also been applied to CDR models for conversational search. In particular, we incorporated a query rewritten module that utilizes T5-Large to deal with conversational queries and relied on pseudo-relevance labels generated using T5-3B and SimANS on the rewritten queries. The experiments revealed that this approach yields state-of-the-art CDR models for domain adaptation. 
 Overall, by making use of real queries and documents of the target domain, the simple labeling approach we have followed, combined with query generation or query rewriting, has proved to be very effective for adapting or further improving a DR or CDR model to new domains.

\nocite{*}
\section{Bibliographical References}\label{sec:reference}

\bibliographystyle{lrec-coling2024-natbib}
\bibliography{lrec-coling2024-example}

\begin{thebibliography}{68}
\expandafter\ifx\csname natexlab\endcsname\relax\def\natexlab#1{#1}\fi

\bibitem[{Blanchard et~al.(2021)Blanchard, Deshmukh, Dogan, Lee, and Scott}]{blanchard2021domain}
Gilles Blanchard, Aniket~Anand Deshmukh, {\"U}run Dogan, Gyemin Lee, and Clayton Scott. 2021.
\newblock Domain generalization by marginal transfer learning.
\newblock \emph{The Journal of Machine Learning Research}, 22(1):46--100.

\bibitem[{Burges(2010)}]{burges2010ranknet}
Christopher J.~C. Burges. 2010.
\newblock From {RankNet} to {LambdaRank} to {LambdaMART}: An overview.
\newblock Technical report, Microsoft Research.

\bibitem[{Carlucci et~al.(2019)Carlucci, D'Innocente, Bucci, Caputo, and Tommasi}]{carlucci2019domain}
Fabio~M Carlucci, Antonio D'Innocente, Silvia Bucci, Barbara Caputo, and Tatiana Tommasi. 2019.
\newblock Domain generalization by solving jigsaw puzzles.
\newblock In \emph{Proceedings of the IEEE/CVF Conference on Computer Vision and Pattern Recognition}, pages 2229--2238.

\bibitem[{Culpepper et~al.(2018)Culpepper, Diaz, and Smucker}]{culpepper2018research}
J.~Shane Culpepper, Fernando Diaz, and Mark~D Smucker. 2018.
\newblock Research frontiers in information retrieval: Report from the third strategic workshop on information retrieval in lorne (swirl 2018).
\newblock In \emph{ACM SIGIR Forum}, volume~52, pages 34--90. ACM New York, NY, USA.

\bibitem[{Dai et~al.(2022)Dai, Zhao, Ma, Luan, Ni, Lu, Bakalov, Guu, Hall, and Chang}]{dai2022promptagator}
Zhuyun Dai, Vincent~Y Zhao, Ji~Ma, Yi~Luan, Jianmo Ni, Jing Lu, Anton Bakalov, Kelvin Guu, Keith Hall, and Ming-Wei Chang. 2022.
\newblock Promptagator: Few-shot dense retrieval from 8 examples.
\newblock In \emph{The Eleventh International Conference on Learning Representations}.

\bibitem[{Dalton et~al.(2020)Dalton, Xiong, and Callan}]{dalton2020trec}
Jeffrey Dalton, Chenyan Xiong, and Jamie Callan. 2020.
\newblock Trec cast 2019: The conversational assistance track overview.
\newblock \emph{arXiv preprint arXiv:2003.13624}.

\bibitem[{Du et~al.(2020)Du, Xu, Xiong, Qiu, Zhen, Snoek, and Shao}]{du2020learning}
Yingjun Du, Jun Xu, Huan Xiong, Qiang Qiu, Xiantong Zhen, Cees~GM Snoek, and Ling Shao. 2020.
\newblock Learning to learn with variational information bottleneck for domain generalization.
\newblock In \emph{European Conference on Computer Vision}, pages 200--216. Springer.

\bibitem[{D’Innocente and Caputo(2018)}]{d2018domain}
Antonio D’Innocente and Barbara Caputo. 2018.
\newblock Domain generalization with domain-specific aggregation modules.
\newblock In \emph{German Conference on Pattern Recognition}, pages 187--198. Springer.

\bibitem[{Elgohary et~al.(2019)Elgohary, Peskov, and Boyd-Graber}]{elgohary2019can}
Ahmed Elgohary, Denis Peskov, and Jordan Boyd-Graber. 2019.
\newblock Can you unpack that? learning to rewrite questions-in-context.
\newblock In \emph{Proceedings of the 2019 Conference on Empirical Methods in Natural Language Processing and the 9th International Joint Conference on Natural Language Processing (EMNLP-IJCNLP)}, pages 5918--5924.

\bibitem[{Ganin et~al.(2016)Ganin, Ustinova, Ajakan, Germain, Larochelle, Laviolette, Marchand, and Lempitsky}]{ganin2016domain}
Yaroslav Ganin, Evgeniya Ustinova, Hana Ajakan, Pascal Germain, Hugo Larochelle, Fran{\c{c}}ois Laviolette, Mario Marchand, and Victor Lempitsky. 2016.
\newblock Domain-adversarial training of neural networks.
\newblock \emph{The journal of machine learning research}, 17(1):2096--2030.

\bibitem[{Gao and Callan(2021)}]{gao2021condenser}
Luyu Gao and Jamie Callan. 2021.
\newblock Condenser: a pre-training architecture for dense retrieval.
\newblock In \emph{Proceedings of the 2021 Conference on Empirical Methods in Natural Language Processing}, pages 981--993.

\bibitem[{Gao and Callan(2022)}]{gao2022unsupervised}
Luyu Gao and Jamie Callan. 2022.
\newblock Unsupervised corpus aware language model pre-training for dense passage retrieval.
\newblock In \emph{Proceedings of the 60th Annual Meeting of the Association for Computational Linguistics (Volume 1: Long Papers)}, pages 2843--2853.

\bibitem[{Gao et~al.(2021)Gao, Yao, and Chen}]{gao2021simcse}
Tianyu Gao, Xingcheng Yao, and Danqi Chen. 2021.
\newblock Simcse: Simple contrastive learning of sentence embeddings.
\newblock In \emph{Proceedings of the 2021 Conference on Empirical Methods in Natural Language Processing}, pages 6894--6910.

\bibitem[{Guo et~al.(2020)Guo, Fan, Pang, Yang, Ai, Zamani, Wu, Croft, and Cheng}]{guo2020deep}
Jiafeng Guo, Yixing Fan, Liang Pang, Liu Yang, Qingyao Ai, Hamed Zamani, Chen Wu, W~Bruce Croft, and Xueqi Cheng. 2020.
\newblock A deep look into neural ranking models for information retrieval.
\newblock \emph{Information Processing \& Management}, 57(6):102067.

\bibitem[{Hofst{\"a}tter et~al.(2020)Hofst{\"a}tter, Althammer, Schr{\"o}der, Sertkan, and Hanbury}]{hofstatter2020improving}
Sebastian Hofst{\"a}tter, Sophia Althammer, Michael Schr{\"o}der, Mete Sertkan, and Allan Hanbury. 2020.
\newblock Improving efficient neural ranking models with cross-architecture knowledge distillation.
\newblock \emph{arXiv preprint arXiv:2010.02666}.

\bibitem[{Jeon et~al.(2021)Jeon, Hong, Lee, Lee, and Byun}]{jeon2021feature}
Seogkyu Jeon, Kibeom Hong, Pilhyeon Lee, Jewook Lee, and Hyeran Byun. 2021.
\newblock Feature stylization and domain-aware contrastive learning for domain generalization.
\newblock In \emph{Proceedings of the 29th ACM International Conference on Multimedia}, pages 22--31.

\bibitem[{Karouzos et~al.(2021)Karouzos, Paraskevopoulos, and Potamianos}]{karouzos2021udalm}
Constantinos Karouzos, Georgios Paraskevopoulos, and Alexandros Potamianos. 2021.
\newblock Udalm: Unsupervised domain adaptation through language modeling.
\newblock In \emph{Proceedings of the 2021 Conference of the North American Chapter of the Association for Computational Linguistics: Human Language Technologies}, pages 2579--2590.

\bibitem[{Karpukhin et~al.(2020)Karpukhin, Oguz, Min, Lewis, Wu, Edunov, Chen, and Yih}]{karpukhin2020dense}
Vladimir Karpukhin, Barlas Oguz, Sewon Min, Patrick Lewis, Ledell Wu, Sergey Edunov, Danqi Chen, and Wen-tau Yih. 2020.
\newblock Dense passage retrieval for open-domain question answering.
\newblock In \emph{Proceedings of the 2020 Conference on Empirical Methods in Natural Language Processing (EMNLP)}, pages 6769--6781.

\bibitem[{Khattab and Zaharia(2020)}]{khattab2020colbert}
Omar Khattab and Matei Zaharia. 2020.
\newblock Colbert: Efficient and effective passage search via contextualized late interaction over bert.
\newblock In \emph{Proceedings of the 43rd International ACM SIGIR conference on research and development in Information Retrieval}, pages 39--48.

\bibitem[{Lee et~al.(2019)Lee, Chang, and Toutanova}]{lee2019latent}
Kenton Lee, Ming-Wei Chang, and Kristina Toutanova. 2019.
\newblock Latent retrieval for weakly supervised open domain question answering.
\newblock In \emph{Proceedings of the 57th Annual Meeting of the Association for Computational Linguistics}, pages 6086--6096.

\bibitem[{Li et~al.(2018)Li, Yang, Song, and Hospedales}]{li2018learning}
Da~Li, Yongxin Yang, Yi-Zhe Song, and Timothy Hospedales. 2018.
\newblock Learning to generalize: Meta-learning for domain generalization.
\newblock In \emph{Proceedings of the AAAI conference on artificial intelligence}, volume~32.

\bibitem[{Li et~al.(2017)Li, Yang, Song, and Hospedales}]{li2017deeper}
Da~Li, Yongxin Yang, Yi-Zhe Song, and Timothy~M Hospedales. 2017.
\newblock Deeper, broader and artier domain generalization.
\newblock In \emph{Proceedings of the IEEE international conference on computer vision}, pages 5542--5550.

\bibitem[{Li and Gaussier(2022)}]{li2022bert}
Minghan Li and Eric Gaussier. 2022.
\newblock Bert-based dense intra-ranking and contextualized late interaction via multi-task learning for long document retrieval.
\newblock In \emph{Proceedings of the 45th International ACM SIGIR Conference on Research and Development in Information Retrieval}, pages 2347--2352.

\bibitem[{Liang et~al.(2020)Liang, Xu, Shakeri, Santos, Nallapati, Huang, and Xiang}]{liang2020embedding}
Davis Liang, Peng Xu, Siamak Shakeri, Cicero Nogueira~dos Santos, Ramesh Nallapati, Zhiheng Huang, and Bing Xiang. 2020.
\newblock Embedding-based zero-shot retrieval through query generation.
\newblock \emph{arXiv preprint arXiv:2009.10270}.

\bibitem[{Lin et~al.(2021)Lin, Yang, and Lin}]{lin2021contextualized}
Sheng-Chieh Lin, Jheng-Hong Yang, and Jimmy Lin. 2021.
\newblock Contextualized query embeddings for conversational search.
\newblock In \emph{Proceedings of the 2021 Conference on Empirical Methods in Natural Language Processing}, pages 1004--1015.

\bibitem[{Liu et~al.(2021)Liu, Sun, Wang, Tang, Li, Qin, Chen, and Liu}]{liu2021learning}
Chang Liu, Xinwei Sun, Jindong Wang, Haoyue Tang, Tao Li, Tao Qin, Wei Chen, and Tie-Yan Liu. 2021.
\newblock Learning causal semantic representation for out-of-distribution prediction.
\newblock \emph{Advances in Neural Information Processing Systems}, 34:6155--6170.

\bibitem[{Loshchilov and Hutter(2017{\natexlab{a}})}]{loshchilov2017decoupled}
Ilya Loshchilov and Frank Hutter. 2017{\natexlab{a}}.
\newblock Decoupled weight decay regularization.
\newblock \emph{arXiv preprint arXiv:1711.05101}.

\bibitem[{Loshchilov and Hutter(2017{\natexlab{b}})}]{loshchilov2017sgdr}
Ilya Loshchilov and Frank Hutter. 2017{\natexlab{b}}.
\newblock \href {https://openreview.net/forum?id=Skq89Scxx} {{SGDR}: Stochastic gradient descent with warm restarts}.
\newblock In \emph{International Conference on Learning Representations}.

\bibitem[{Ma et~al.(2021)Ma, Korotkov, Yang, Hall, and McDonald}]{ma2021zero}
Ji~Ma, Ivan Korotkov, Yinfei Yang, Keith Hall, and Ryan McDonald. 2021.
\newblock Zero-shot neural passage retrieval via domain-targeted synthetic question generation.
\newblock In \emph{Proceedings of the 16th Conference of the European Chapter of the Association for Computational Linguistics: Main Volume}, pages 1075--1088.

\bibitem[{Maia et~al.(2018)Maia, Handschuh, Freitas, Davis, McDermott, Zarrouk, and Balahur}]{fiqa-2018}
Macedo Maia, Siegfried Handschuh, Andr\'{e} Freitas, Brian Davis, Ross McDermott, Manel Zarrouk, and Alexandra Balahur. 2018.
\newblock \href {https://doi.org/10.1145/3184558.3192301} {Www'18 open challenge: Financial opinion mining and question answering}.
\newblock In \emph{Companion Proceedings of the The Web Conference 2018}, WWW '18, page 1941–1942, Republic and Canton of Geneva, CHE. International World Wide Web Conferences Steering Committee.

\bibitem[{Mancini et~al.(2018)Mancini, Bulo, Caputo, and Ricci}]{mancini2018best}
Massimiliano Mancini, Samuel~Rota Bulo, Barbara Caputo, and Elisa Ricci. 2018.
\newblock Best sources forward: domain generalization through source-specific nets.
\newblock In \emph{2018 25th IEEE international conference on image processing (ICIP)}, pages 1353--1357. IEEE.

\bibitem[{Mao et~al.(2022)Mao, Dou, Qian, Mo, Cheng, and Cao}]{mao2022convtrans}
Kelong Mao, Zhicheng Dou, Hongjin Qian, Fengran Mo, Xiaohua Cheng, and Zhao Cao. 2022.
\newblock Convtrans: Transforming web search sessions for conversational dense retrieval.
\newblock In \emph{Proceedings of the 2022 Conference on Empirical Methods in Natural Language Processing}, pages 2935--2946.

\bibitem[{Mele et~al.(2020)Mele, Muntean, Nardini, Perego, Tonellotto, and Frieder}]{mele2020topic}
Ida Mele, Cristina~Ioana Muntean, Franco~Maria Nardini, Raffaele Perego, Nicola Tonellotto, and Ophir Frieder. 2020.
\newblock Topic propagation in conversational search.
\newblock In \emph{Proceedings of the 43rd International ACM SIGIR conference on research and development in Information Retrieval}, pages 2057--2060.

\bibitem[{Micikevicius et~al.(2018)Micikevicius, Narang, Alben, Diamos, Elsen, Garcia, Ginsburg, Houston, Kuchaiev, Venkatesh et~al.}]{micikevicius2018mixed}
Paulius Micikevicius, Sharan Narang, Jonah Alben, Gregory Diamos, Erich Elsen, David Garcia, Boris Ginsburg, Michael Houston, Oleksii Kuchaiev, Ganesh Venkatesh, et~al. 2018.
\newblock Mixed precision training.
\newblock In \emph{International Conference on Learning Representations}.

\bibitem[{Mokrii et~al.(2021)Mokrii, Boytsov, and Braslavski}]{mokrii2021systematic}
Iurii Mokrii, Leonid Boytsov, and Pavel Braslavski. 2021.
\newblock A systematic evaluation of transfer learning and pseudo-labeling with bert-based ranking models.
\newblock In \emph{Proceedings of the 44th International ACM SIGIR Conference on Research and Development in Information Retrieval}, pages 2081--2085.

\bibitem[{Motiian et~al.(2017)Motiian, Piccirilli, Adjeroh, and Doretto}]{motiian2017unified}
Saeid Motiian, Marco Piccirilli, Donald~A Adjeroh, and Gianfranco Doretto. 2017.
\newblock Unified deep supervised domain adaptation and generalization.
\newblock In \emph{Proceedings of the IEEE international conference on computer vision}, pages 5715--5725.

\bibitem[{Nam et~al.(2021)Nam, Lee, Park, Yoon, and Yoo}]{nam2021reducing}
Hyeonseob Nam, HyunJae Lee, Jongchan Park, Wonjun Yoon, and Donggeun Yoo. 2021.
\newblock Reducing domain gap by reducing style bias.
\newblock In \emph{Proceedings of the IEEE/CVF Conference on Computer Vision and Pattern Recognition}, pages 8690--8699.

\bibitem[{Nguyen et~al.(2016)Nguyen, Rosenberg, Song, Gao, Tiwary, Majumder, and Deng}]{bajaj2018ms}
Tri Nguyen, Mir Rosenberg, Xia Song, Jianfeng Gao, Saurabh Tiwary, Rangan Majumder, and Li~Deng. 2016.
\newblock {MS} {MARCO:} {A} human generated machine reading comprehension dataset.
\newblock In \emph{Proceedings of the Workshop on Cognitive Computation: Integrating neural and symbolic approaches 2016 co-located with the 30th Annual Conference on Neural Information Processing Systems {(NIPS} 2016), Barcelona, Spain, December 9, 2016}, volume 1773 of \emph{{CEUR} Workshop Proceedings}. CEUR-WS.org.

\bibitem[{Nogueira et~al.(2020)Nogueira, Jiang, Pradeep, and Lin}]{nogueira2020document}
Rodrigo Nogueira, Zhiying Jiang, Ronak Pradeep, and Jimmy Lin. 2020.
\newblock Document ranking with a pretrained sequence-to-sequence model.
\newblock In \emph{Findings of the Association for Computational Linguistics: EMNLP 2020}, pages 708--718.

\bibitem[{Papineni et~al.(2002)Papineni, Roukos, Ward, and Zhu}]{papineni2002bleu}
Kishore Papineni, Salim Roukos, Todd Ward, and Wei-Jing Zhu. 2002.
\newblock Bleu: a method for automatic evaluation of machine translation.
\newblock In \emph{Proceedings of the 40th annual meeting of the Association for Computational Linguistics}, pages 311--318.

\bibitem[{Prakash et~al.(2019)Prakash, Boochoon, Brophy, Acuna, Cameracci, State, Shapira, and Birchfield}]{prakash2019structured}
Aayush Prakash, Shaad Boochoon, Mark Brophy, David Acuna, Eric Cameracci, Gavriel State, Omer Shapira, and Stan Birchfield. 2019.
\newblock Structured domain randomization: Bridging the reality gap by context-aware synthetic data.
\newblock In \emph{2019 International Conference on Robotics and Automation (ICRA)}, pages 7249--7255. IEEE.

\bibitem[{Qiao et~al.(2020)Qiao, Zhao, and Peng}]{qiao2020learning}
Fengchun Qiao, Long Zhao, and Xi~Peng. 2020.
\newblock Learning to learn single domain generalization.
\newblock In \emph{Proceedings of the IEEE/CVF Conference on Computer Vision and Pattern Recognition}, pages 12556--12565.

\bibitem[{Qu et~al.(2021)Qu, Ding, Liu, Liu, Ren, Zhao, Dong, Wu, and Wang}]{qu2021rocketqa}
Yingqi Qu, Yuchen Ding, Jing Liu, Kai Liu, Ruiyang Ren, Wayne~Xin Zhao, Daxiang Dong, Hua Wu, and Haifeng Wang. 2021.
\newblock Rocketqa: An optimized training approach to dense passage retrieval for open-domain question answering.
\newblock In \emph{Proceedings of the 2021 Conference of the North American Chapter of the Association for Computational Linguistics: Human Language Technologies}, pages 5835--5847.

\bibitem[{Raffel et~al.(2020)Raffel, Shazeer, Roberts, Lee, Narang, Matena, Zhou, Li, Liu et~al.}]{raffel2020exploring}
Colin Raffel, Noam Shazeer, Adam Roberts, Katherine Lee, Sharan Narang, Michael Matena, Yanqi Zhou, Wei Li, Peter~J Liu, et~al. 2020.
\newblock Exploring the limits of transfer learning with a unified text-to-text transformer.
\newblock \emph{J. Mach. Learn. Res.}, 21(140):1--67.

\bibitem[{Rahman et~al.(2019)Rahman, Fookes, Baktashmotlagh, and Sridharan}]{rahman2019multi}
Mohammad~Mahfujur Rahman, Clinton Fookes, Mahsa Baktashmotlagh, and Sridha Sridharan. 2019.
\newblock Multi-component image translation for deep domain generalization.
\newblock In \emph{2019 IEEE Winter Conference on Applications of Computer Vision (WACV)}, pages 579--588. IEEE.

\bibitem[{Reimers and Gurevych(2019)}]{reimers2019sentence}
Nils Reimers and Iryna Gurevych. 2019.
\newblock Sentence-bert: Sentence embeddings using siamese bert-networks.
\newblock In \emph{Proceedings of the 2019 Conference on Empirical Methods in Natural Language Processing and the 9th International Joint Conference on Natural Language Processing (EMNLP-IJCNLP)}, pages 3982--3992.

\bibitem[{Ren et~al.(2018)Ren, Ni, Malik, and Ke}]{ren2018conversational}
Gary Ren, Xiaochuan Ni, Manish Malik, and Qifa Ke. 2018.
\newblock Conversational query understanding using sequence to sequence modeling.
\newblock In \emph{Proceedings of the 2018 World Wide Web Conference}, pages 1715--1724.

\bibitem[{Robertson and Zaragoza(2009)}]{RobertsonZ09}
Stephen~E. Robertson and Hugo Zaragoza. 2009.
\newblock The probabilistic relevance framework: {BM25} and beyond.
\newblock \emph{Found. Trends Inf. Retr.}, 3(4):333--389.

\bibitem[{Sanh et~al.(2019)Sanh, Debut, Chaumond, and Wolf}]{DBLP:journals/corr/abs-1910-01108}
Victor Sanh, Lysandre Debut, Julien Chaumond, and Thomas Wolf. 2019.
\newblock Distilbert, a distilled version of {BERT:} smaller, faster, cheaper and lighter.
\newblock \emph{arXiv preprint arXiv:1910.01108}.

\bibitem[{Shankar et~al.(2018)Shankar, Piratla, Chakrabarti, Chaudhuri, Jyothi, and Sarawagi}]{shankar2018generalizing}
Shiv Shankar, Vihari Piratla, Soumen Chakrabarti, Siddhartha Chaudhuri, Preethi Jyothi, and Sunita Sarawagi. 2018.
\newblock Generalizing across domains via cross-gradient training.
\newblock In \emph{International Conference on Learning Representations}.

\bibitem[{Sun et~al.(2021)Sun, Qian, Liu, Xiong, Zhang, Bao, Liu, and Bennett}]{sun2021few}
Si~Sun, Yingzhuo Qian, Zhenghao Liu, Chenyan Xiong, Kaitao Zhang, Jie Bao, Zhiyuan Liu, and Paul Bennett. 2021.
\newblock Few-shot text ranking with meta adapted synthetic weak supervision.
\newblock In \emph{Proceedings of the 59th Annual Meeting of the Association for Computational Linguistics and the 11th International Joint Conference on Natural Language Processing (Volume 1: Long Papers)}, pages 5030--5043.

\bibitem[{Thakur et~al.(2021)Thakur, Reimers, R{\"u}ckl{\'e}, Srivastava, and Gurevych}]{thakur2021beir}
Nandan Thakur, Nils Reimers, Andreas R{\"u}ckl{\'e}, Abhishek Srivastava, and Iryna Gurevych. 2021.
\newblock {BEIR}: A heterogeneous benchmark for zero-shot evaluation of information retrieval models.
\newblock In \emph{Thirty-fifth Conference on Neural Information Processing Systems Datasets and Benchmarks Track (Round 2)}.

\bibitem[{Tobin et~al.(2017)Tobin, Fong, Ray, Schneider, Zaremba, and Abbeel}]{tobin2017domain}
Josh Tobin, Rachel Fong, Alex Ray, Jonas Schneider, Wojciech Zaremba, and Pieter Abbeel. 2017.
\newblock Domain randomization for transferring deep neural networks from simulation to the real world.
\newblock In \emph{2017 IEEE/RSJ international conference on intelligent robots and systems (IROS)}, pages 23--30. IEEE.

\bibitem[{Tsatsaronis et~al.(2015)Tsatsaronis, Balikas, Malakasiotis, Partalas, Zschunke, Alvers, Weissenborn, Krithara, Petridis, Polychronopoulos et~al.}]{bioasq-2015}
George Tsatsaronis, Georgios Balikas, Prodromos Malakasiotis, Ioannis Partalas, Matthias Zschunke, Michael~R Alvers, Dirk Weissenborn, Anastasia Krithara, Sergios Petridis, Dimitris Polychronopoulos, et~al. 2015.
\newblock An overview of the bioasq large-scale biomedical semantic indexing and question answering competition.
\newblock \emph{BMC bioinformatics}, 16(1):138.

\bibitem[{Vakulenko et~al.(2021)Vakulenko, Longpre, Tu, and Anantha}]{vakulenko2021question}
Svitlana Vakulenko, Shayne Longpre, Zhucheng Tu, and Raviteja Anantha. 2021.
\newblock Question rewriting for conversational question answering.
\newblock In \emph{Proceedings of the 14th ACM international conference on web search and data mining}, pages 355--363.

\bibitem[{Volpi et~al.(2018)Volpi, Namkoong, Sener, Duchi, Murino, and Savarese}]{volpi2018generalizing}
Riccardo Volpi, Hongseok Namkoong, Ozan Sener, John~C Duchi, Vittorio Murino, and Silvio Savarese. 2018.
\newblock Generalizing to unseen domains via adversarial data augmentation.
\newblock \emph{Advances in neural information processing systems}, 31.

\bibitem[{Voorhees(2005)}]{robust04-2005}
Ellen Voorhees. 2005.
\newblock Overview of the trec 2004 robust retrieval track.
\newblock Special Publication (NIST SP), National Institute of Standards and Technology, Gaithersburg, MD.

\bibitem[{Wang et~al.(2022{\natexlab{a}})Wang, Lan, Liu, Ouyang, Qin, Lu, Chen, Zeng, and Yu}]{wang2022generalizing}
Jindong Wang, Cuiling Lan, Chang Liu, Yidong Ouyang, Tao Qin, Wang Lu, Yiqiang Chen, Wenjun Zeng, and Philip Yu. 2022{\natexlab{a}}.
\newblock Generalizing to unseen domains: A survey on domain generalization.
\newblock \emph{IEEE Transactions on Knowledge and Data Engineering}.

\bibitem[{Wang et~al.(2021)Wang, Reimers, and Gurevych}]{wang2021tsdae}
Kexin Wang, Nils Reimers, and Iryna Gurevych. 2021.
\newblock Tsdae: Using transformer-based sequential denoising auto-encoderfor unsupervised sentence embedding learning.
\newblock In \emph{Findings of the Association for Computational Linguistics: EMNLP 2021}, pages 671--688.

\bibitem[{Wang et~al.(2022{\natexlab{b}})Wang, Thakur, Reimers, and Gurevych}]{wang-etal-2022-gpl}
Kexin Wang, Nandan Thakur, Nils Reimers, and Iryna Gurevych. 2022{\natexlab{b}}.
\newblock \href {https://doi.org/10.18653/v1/2022.naacl-main.168} {{GPL}: Generative pseudo labeling for unsupervised domain adaptation of dense retrieval}.
\newblock In \emph{Proceedings of the 2022 Conference of the North American Chapter of the Association for Computational Linguistics: Human Language Technologies}, pages 2345--2360, Seattle, United States. Association for Computational Linguistics.

\bibitem[{Wang and Deng(2018)}]{wang2018DA}
M.~Wang and W.~Deng. 2018.
\newblock Deep visual domain adaptation: a survey.
\newblock \emph{Neurocomputing}.

\bibitem[{Xin et~al.(2022)Xin, Xiong, Srinivasan, Sharma, Jose, and Bennett}]{xin2022zero}
Ji~Xin, Chenyan Xiong, Ashwin Srinivasan, Ankita Sharma, Damien Jose, and Paul Bennett. 2022.
\newblock Zero-shot dense retrieval with momentum adversarial domain invariant representations.
\newblock In \emph{Findings of the Association for Computational Linguistics: ACL 2022}, pages 4008--4020.

\bibitem[{Xiong et~al.(2020)Xiong, Xiong, Li, Tang, Liu, Bennett, Ahmed, and Overwijk}]{xiong2020approximate}
Lee Xiong, Chenyan Xiong, Ye~Li, Kwok-Fung Tang, Jialin Liu, Paul~N Bennett, Junaid Ahmed, and Arnold Overwijk. 2020.
\newblock Approximate nearest neighbor negative contrastive learning for dense text retrieval.
\newblock In \emph{International Conference on Learning Representations}.

\bibitem[{Yang et~al.(2018)Yang, Fang, and Lin}]{yang2018anserini}
Peilin Yang, Hui Fang, and Jimmy Lin. 2018.
\newblock Anserini: Reproducible ranking baselines using lucene.
\newblock \emph{Journal of Data and Information Quality (JDIQ)}, 10(4):1--20.

\bibitem[{Yu et~al.(2021)Yu, Liu, Xiong, Feng, and Liu}]{yu2021few}
Shi Yu, Zhenghao Liu, Chenyan Xiong, Tao Feng, and Zhiyuan Liu. 2021.
\newblock Few-shot conversational dense retrieval.
\newblock In \emph{Proceedings of the 44th International ACM SIGIR Conference on research and development in information retrieval}, pages 829--838.

\bibitem[{Zamani et~al.(2023)Zamani, Trippas, Dalton, and Radlinski}]{zamani2023conversational}
Hamed Zamani, Johanne~R. Trippas, Jeff Dalton, and Filip Radlinski. 2023.
\newblock \href {http://arxiv.org/abs/2201.08808} {Conversational information seeking}.

\bibitem[{Zhang et~al.(2018)Zhang, Cisse, Dauphin, and Lopez-Paz}]{zhang2018mixup}
Hongyi Zhang, Moustapha Cisse, Yann~N Dauphin, and David Lopez-Paz. 2018.
\newblock mixup: Beyond empirical risk minimization.
\newblock In \emph{International Conference on Learning Representations}.

\bibitem[{Zhou et~al.(2022)Zhou, Gong, Liu, Zhao, Shen, Dong, Lu, Majumder, Wen, and Duan}]{zhou2022simans}
Kun Zhou, Yeyun Gong, Xiao Liu, Wayne~Xin Zhao, Yelong Shen, Anlei Dong, Jingwen Lu, Rangan Majumder, Ji-rong Wen, and Nan Duan. 2022.
\newblock \href {https://aclanthology.org/2022.emnlp-industry.56} {{S}im{ANS}: Simple ambiguous negatives sampling for dense text retrieval}.
\newblock In \emph{Proceedings of the 2022 Conference on Empirical Methods in Natural Language Processing: Industry Track}, pages 548--559, Abu Dhabi, UAE. Association for Computational Linguistics.

\end{thebibliography}

\end{document}